\pdfoutput=1

\documentclass[11pt,twoside,a4paper,cmspaper,final,collab]{cms-tdr}
\def\svnVersion{386448}\def\cmsCernNoTag{CERN-EP/2016-096}\def\cmsCernDate{\today}\def\cmsMessage{Published in Physical Review Letters as \href{http://dx.doi.org/10.1103/PhysRevLett.118.021802}{\doi{10.1103/PhysRevLett.118.021802.}}}

\begin{document}\cmsNoteHeader{SUS-14-019}

\hyphenation{had-ron-i-za-tion}
\hyphenation{cal-or-i-me-ter}
\hyphenation{de-vices}
\RCS$Revision: 374296 $
\RCS$HeadURL: svn+ssh://svn.cern.ch/reps/tdr2/papers/SUS-14-019/trunk/SUS-14-019.tex $
\RCS$Id: SUS-14-019.tex 374296 2016-11-18 21:35:27Z adelanno $
\newlength\cmsFigWidth
\ifthenelse{\boolean{cms@external}}{\setlength\cmsFigWidth{0.85\columnwidth}}{\setlength\cmsFigWidth{0.4\textwidth}}
\ifthenelse{\boolean{cms@external}}{\providecommand{\cmsLeft}{top\xspace}}{\providecommand{\cmsLeft}{left\xspace}}
\ifthenelse{\boolean{cms@external}}{\providecommand{\cmsRight}{bottom\xspace}}{\providecommand{\cmsRight}{right\xspace}}
\providecommand{\PV}{\cmsSymbolFace{V}\xspace}
\ifthenelse{\boolean{cms@external}}{\providecommand{\NA}{\ensuremath{\cdots}\xspace}}{\providecommand{\NA}{\ensuremath{\text{---}}\xspace}}
\newcommand{\mpt}{\ensuremath{\pt^{\text{miss}}}\xspace}
\newcommand{\CLs}{\ensuremath{\mathrm{CL}_\mathrm{s}}\xspace}
\ifthenelse{\boolean{cms@external}}{\providecommand{\CL}{C.L.\xspace}}{\providecommand{\CL}{CL\xspace}}
\ifthenelse{\boolean{cms@external}}{\newcommand{\cmsTable}[1]{#1}}{\newcommand{\cmsTable}[1]{\resizebox{\textwidth}{!}{#1}}}

\cmsNoteHeader{SUS-14-019}
\title{Search for dark matter and supersymmetry with a compressed mass spectrum in the vector boson fusion topology in proton-proton collisions at \texorpdfstring{$\sqrt{s} = 8$\TeV}{sqrt(s) = 8 TeV}}

\date{\today}

\abstract{
A first search for pair production of dark matter candidates through vector boson fusion in proton-proton collisions at $\sqrt{s} = 8$\TeV is performed with the CMS detector.
The vector boson fusion topology enhances missing transverse momentum, providing a way to probe supersymmetry even in the case of a compressed mass spectrum.
The data sample corresponds to an integrated luminosity of 18.5\fbinv recorded by the CMS experiment.
The observed dijet mass spectrum is consistent with the standard model expectation.
In an effective field theory, dark matter masses are explored as a function of contact interaction strength.
The most stringent limit on bottom squark production with mass below 315\GeV is also reported, assuming a 5\GeV mass difference with respect to the lightest neutralino.
}

\hypersetup{
pdfauthor={CMS Collaboration},
pdftitle={Search for dark matter and supersymmetry with a compressed mass spectrum in the vector boson fusion topology in proton-proton collisions at sqrt(s) = 8 TeV},
pdfsubject={CMS},
pdfkeywords={CMS, physics, SUSY, VBF}}

\maketitle

Cosmological measurements indicate that dark matter (DM) constitutes 85\% of all matter in the universe~\cite{Ade:2013zuv}.
The identity of DM is one of the most fundamental open questions in both particle physics and cosmology.
Many extensions of the standard model (SM) predict a DM candidate in the form of a weakly interacting massive particle (WIMP) at the electroweak symmetry breaking scale.

Previously, searches for WIMP DM at the CERN LHC have been performed in the context of complete renormalizable theories, such as supersymmetry (SUSY).
For example, many searches for the lightest SUSY particle (LSP) in R-parity conserving SUSY~\cite{Martin:1997ns,Farrar:1978xj} rely on production through decays of heavier particles (\eg squarks) accessible at the LHC that give
rise to signatures with energetic leptons, photons, and/or jets. Such searches~\cite{Khachatryan:2014qwa,Khachatryan:2015pot,Aad:2014nua,Agashe:2014kda} have limited sensitivity in scenarios with a compressed mass spectrum, which
result in visible particles with too little energy to be detected efficiently.

\begin{figure*}[htb]
\hspace*{\fill}
\includegraphics[width=0.3\textwidth]{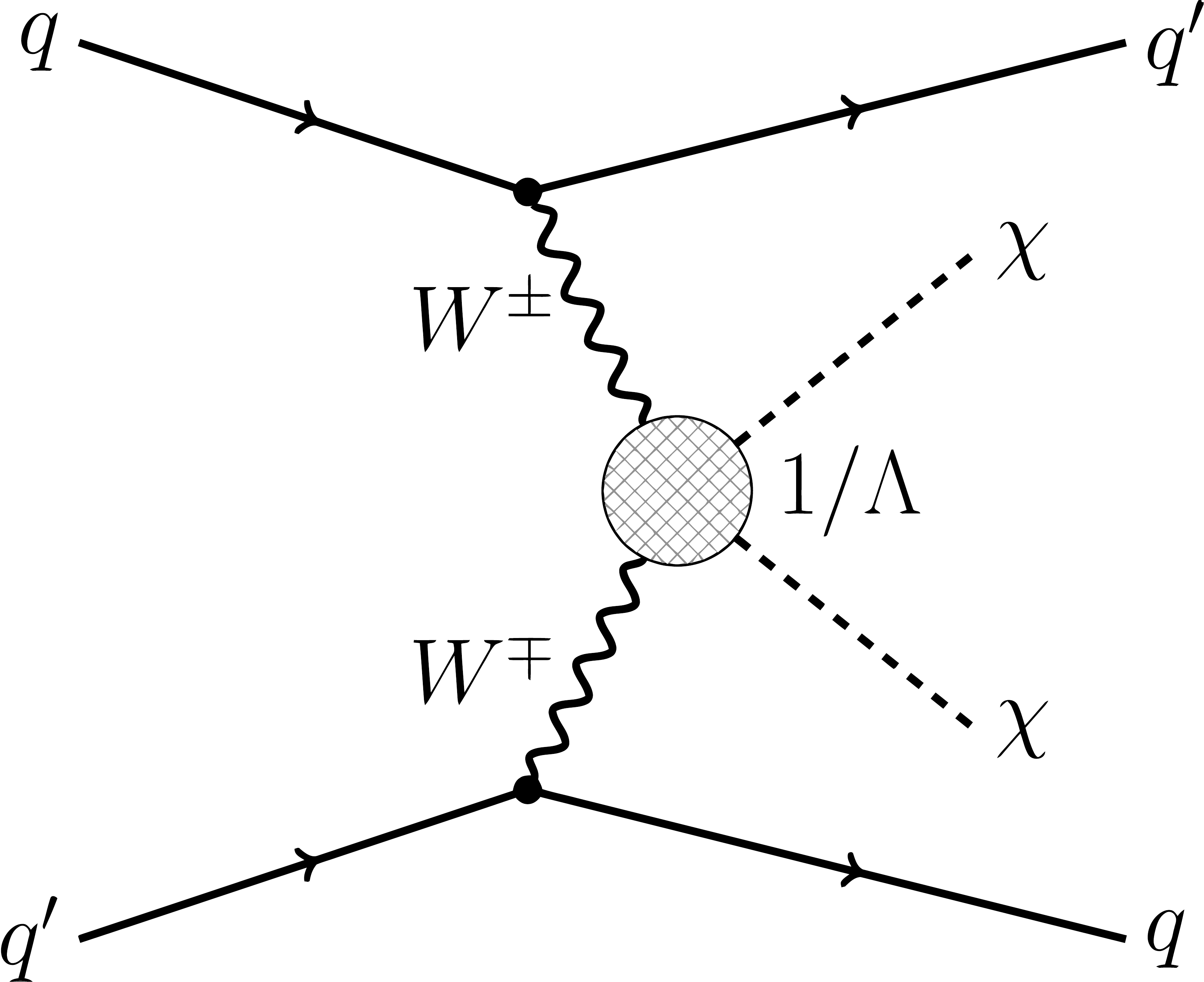}
\hspace*{\fill}
\includegraphics[width=0.3\textwidth]{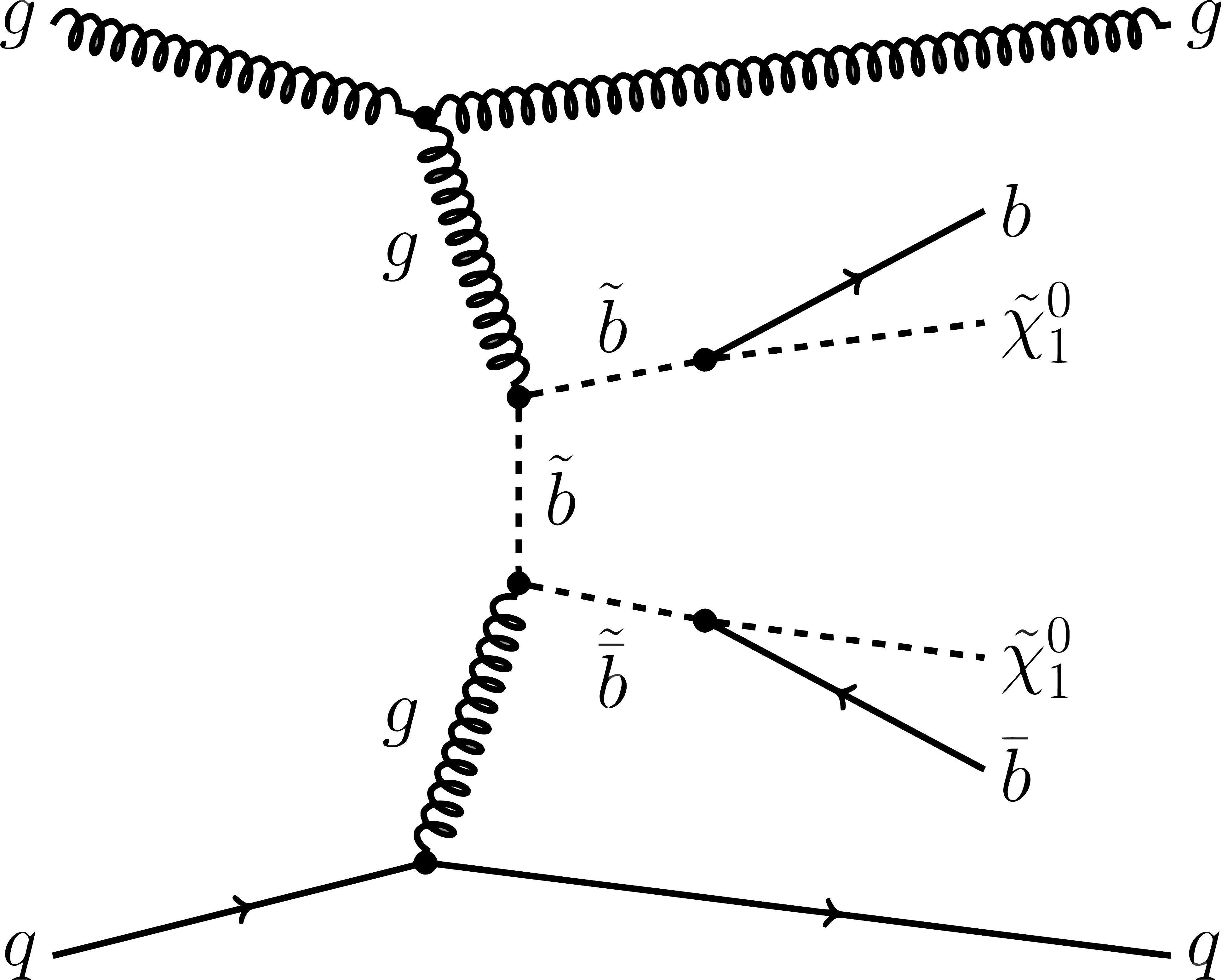}
\hspace*{\fill}
\caption{
        Feynman diagrams for dark matter pair production in a vector boson fusion process (left) and for bottom squark pair production (right). Given a nearly degenerate bottom squark and LSP, the final state b-quarks are too soft to be observed.
}
      \label{fig:feynDiagram}
\end{figure*}

This Letter describes the first search for direct pair production of DM through pure electroweak vector boson fusion (VBF) processes at a hadron collider.
The VBF production mechanism provides a probe of DM that is agnostic to the accessibility of heavier colored or electroweak sectors.
In order to study DM-SM interactions with minimal assumptions, we consider an effective field theory (EFT) approach, which provides complementary information to other DM searches~\cite{monoH1, monoH2, monoV, monoZ}.
The benchmark model used assumes the DM particle to be a Dirac fermion and its interaction with the electroweak gauge bosons to be mediated by a heavy particle (dimension 5a operator as in Ref.~\cite{Cotta:2012nj}).
The EFT framework is examined with a contact interaction of scale $\Lambda = \mathcal{M}/g_{\text{eff}} = \mathcal{M}/\sqrt{g_{\chi}g_\mathrm{V}}$, where $\mathcal{M}$ is the mass of the heavy mediator, $g_{\chi}$ is its coupling to the DM particle, and $g_\mathrm{V}$ is its coupling to vector bosons $\PV=\gamma$, $\Z$, or $\PW$ (Figure~\ref{fig:feynDiagram}, left).

The EFT benchmark model can be used to compare the results in this Letter to other analyses considering V-V-DM-DM contact interactions, but it cannot be directly
compared to searches which probe quark-DM interactions (e.g. in the
monojet topology~\cite{Khachatryan:2015wza, 2015EPJC...75..299A, 2015EPJC...75..235K}).
To demonstrate the effectiveness of this VBF analysis strategy relative to the monojet searches,
we consider as a benchmark the strong production of squarks, which can satisfy the VBF selection
when produced in association with two jets arising from initial-state radiation.
Under the assumption that the squark and the LSP are nearly mass degenerate, the jets produced in the squark decays are typically too soft to be observed.
Here we consider bottom squarks (Figure~\ref{fig:feynDiagram} right) and assume a 5\GeV mass difference with the LSP, where the monojet analyses by ATLAS and CMS~\cite{Khachatryan:2015wza, 2015EPJC...75..299A, 2015EPJC...75..235K} exclude masses below $\approx$250\GeV, but the analysis is applicable to all generations of squarks.

The analysis is performed using data collected with the CMS experiment at the LHC in proton-proton ($\Pp\Pp$) collisions at a center-of-mass energy of 8\TeV.
The data sample corresponds to an integrated luminosity of 18.5\fbinv.
The VBF topology is characterized by the presence of two forward jets (\ie jets near the beam axis) in opposite hemispheres, leading to a large dijet invariant
mass~\cite{Dutta:2012xe,Delannoy:2013ata,SUS-14-005,PhysRevD.93.052002,Chatrchyan:2014tja,Aad:2015txa}.
The two jets boost the decay products of new particles, similar to requiring a jet from initial state radiation, which aids event selection and enhances rejection of multijet background.
We analyze the dijet mass spectrum to search for new physics in events consistent with the VBF topology and with missing transverse momentum ($\mpt$).

The central feature of the CMS apparatus~\cite{Chatrchyan:2008zzk} is a superconducting solenoid of 6\unit{m} internal diameter,
providing a magnetic field of 3.8\unit{T}.
Within the solenoid volume are a silicon pixel and strip tracker,
a lead tungstate crystal electromagnetic calorimeter (ECAL),
and a brass and scintillator hadron calorimeter,
each composed of a barrel and two endcap sections.
Forward calorimeters extend the pseudorapidity ($\eta$) coverage
provided by the barrel and endcap detectors up to $\abs{\eta}<5.2$.
Muons are measured in gas-ionization detectors embedded
in the steel flux-return yoke outside the solenoid.
A more detailed description of the CMS detector,
together with a definition of the coordinate system used (including the azimuthal angle $\phi$) and the relevant kinematic variables
can be found in Ref.~\cite{Chatrchyan:2008zzk}.

The data sample was collected using an online event selection requiring events with $\mpt > 65$\GeV and at least two jets with $\pt > 35$\GeV, with a VBF topology.
This online selection has an efficiency of more than 98\% for the analysis.

For the offline analysis, the events are reconstructed from particle candidates found by the particle-flow (PF) algorithm~\cite{CMS-PAS-PFT-09-001, CMS-PAS-PFT-10-001},
which uses reconstructed objects in an event to build candidate muons, electrons, photons, and charged and neutral hadrons.
The anti-\kt algorithm~\cite{2008JHEP...04..063C} with a distance parameter of 0.5 is used for jet clustering.
Jets are required to pass
identification criteria designed to reject particles from other
interactions in the same bunch crossing (pileup) and spurious
energy measurements in the calorimeters. For jets with $\pt >
30$\GeV and $\abs{\eta}< 2.5$ ($> 2.5$), the identification
efficiency is about 99\% (95\%) with 90\textendash95\% (60\%) of
pileup jets rejected~\cite{CMS:2013wea}.
Jets originating from the hadronization of bottom quarks are tagged using the combined secondary vertex
algorithm~\cite{Chatrchyan:2012jua,CMS:2013vea}. For \PQb-tagged jets
with $\pt > 20$\GeV the identification efficiency is $\approx$85\% with a $\approx$10\% (20\%) misidentification probability for
light quarks and gluons (charm quarks)~\cite{CMS:2013vea}.
The electron momentum is estimated by combining the energy measurement in the ECAL with the momentum measurement in the tracker~\cite{Khachatryan:2015hwa}.
Muons are identified as a track in the central tracker consistent with either a track or several measurements in the muon system, associated with an energy deficit in the calorimeters~\cite{Chatrchyan:2012xi}.
Taus are reconstructed using the hadron plus strips algorithm~\cite{Chatrchyan:2012zz}.

We require exactly two jets with $\pt > 50$\GeV and $\abs{\eta} < 5$ in a VBF configuration, which corresponds to jets in opposite hemispheres ($\eta_1\, \eta_2 < 0$), with large separation in pseudorapidity ($\abs{\Delta\eta}>4.2$), and large dijet mass ($m_\mathrm{jj} > 750$\GeV).
Events with additional jets of $\pt> 30$\GeV (jet veto) or \PQb-tagged jets of $\pt> 20$\GeV are rejected.
Since there are no bottom quarks in (Figure~\ref{fig:feynDiagram} (left) and the bottom quarks in (Figure~\ref{fig:feynDiagram} (right) are too soft to identify efficiently, the rejection of events which
contain a b-tagged jet with $\pt > 20$\GeV is optimized to maintain high
signal efficiency while reducing $\ttbar$ and single-top backgrounds to negligible levels.
Similarly, events with isolated leptons of $\pt> 10$\GeV
($>$15\GeV for tau leptons) and $\abs{\eta}< 2.5$ are
rejected.
For electrons and muons, we define the isolation variable as the \pt sum of the reconstructed PF charged and neutral particles within
a cone of radius $\Delta R = \sqrt{\smash[b]{(\Delta \eta)^2 + (\Delta \phi)^2}}
= 0.3$ centered around the electron or muon track.
We require that this isolation variable divided by the lepton's \pt be less than 0.20.
Isolation for
tau candidates is imposed by applying a dedicated multivariate
discriminator which combines the surrounding energy deposits with
the median energy density flow in the event.
The analysis selects events with $\mpt> 250$\GeV.
To reduce contributions from jet mismeasurements, an azimuthal separation between the subleading jet and the direction
of the missing transverse momentum vector $\abs{ \Delta \phi ( \ptvecmiss, \text{jet}_2 )} > 0.5$ is required.
This set of requirements defines the signal region.

After this selection the main SM contributions are from the production
of $\Z (\to \PGn\PAGn)$ + jets and $\PW (\to \ell \nu)$ + jets (where $\ell=\Pe,\PGm,\PGt$),
with smaller contributions from QCD multijet, $\ttbar$,
and diboson production.
The $\Z (\to \PGn\PAGn)$ + jets background has the same topology as the DM or LSP
signals, and is therefore mostly irreducible.
Due to the contribution to $\mpt$ from neutrinos, $\PW (\to \ell \nu)$ + jets events can enter the signal region if the accompanying charged lepton fails the lepton veto criteria.

Background samples for $\Z (\to \PGn\PAGn)$ + jets, $\PW (\to \ell \nu)$
+ jets, $\ttbar$, and diboson production are
generated with \MADGRAPH (v5.1.3)~\cite{Alwall:2014hca}. Events
with a Higgs boson produced through VBF are generated with \POWHEG
(v1.0r1380)~\cite{Nason:2009ai, Alioli:2010xd}. Signal samples, DM
pair production ($\chi\chi \mathrm{jj}$) and bottom squark pair production
($\PSQb\PASQb\mathrm{jj}$), are generated
with \MADGRAPH (v5.1.5). The momentum distribution of the partons
is taken from {CTEQ6L1} (\MADGRAPH) and {CTEQ6M} (\POWHEG)~\cite{Pumplin:2002vw},
except for the VBF Higgs boson samples where CT10~\cite{Lai:2010vv} is
used. The parton showering, fragmentation, hadronization, and various decays are performed with \PYTHIA (v6.4.22)~\cite{Sjostrand:2006za}.
For background samples, the response of the CMS apparatus is simulated
using \GEANTfour (v9.4p03)~\cite{Agostinelli:2002hh}, while for
the signal samples a fast simulation program~\cite{Abdullin:2011zz}
is used.
The signal acceptance and dijet mass distribution are cross
checked with the \GEANTfour-based simulation and the acceptance
is corrected for the small differences ($<$5\%) observed.
To simulate the effect of pileup, additional pp collisions with the multiplicity distribution matching that in data are superimposed on the hard-scattering event.
Event yields
are normalized to the integrated luminosity of the collision data
using next-to-next-to-leading order cross section calculations,
except in the case of signal samples for which next-to-leading order
($\PSQb\PASQb\mathrm{jj}$)~\cite{Beenakker:1997ut}
 and leading order ($\chi\chi\mathrm{j}$)
cross sections~\cite{Alwall:2014hca} are used.

The strategy for the background estimation is to use Monte Carlo (MC) simulations to model the $\mpt$ distributions, and jet and lepton vetoes.
The background yields predicted by the MC are corrected for observed differences with respect to the data in control regions, and scaled to the fraction of events passing the VBF topology selection, derived from data.
The modeling of the dijet mass distribution is checked in the control regions.
For the $\Z (\to \PGn\PAGn)$ + jets background we use three control regions to
verify the MC simulation, estimate acceptance corrections used to scale the MC yields, and
measure the fraction of events passing the VBF topology selection.
The control regions are defined by treating muons as neutrinos in the $\Z \to \PGmp \PGmm$ decay mode.
The first control region (CR$_{\mathrm{Z1}}$) is a $\Z (\to \PGmp
\PGmm)$ + two jets sample used to validate modeling of geometric
and kinematic acceptance of leptons.
We find a data-to-MC correction of $0.98 \pm 0.01\stat$.
For the CR$_{\mathrm{Z2}}$ control region, which is a subset of CR$_{\mathrm{Z1}}$, we treat the
two muons as neutrinos, subtract the muon \pt vectors from $\ptvecmiss$,
and require $\mpt> 250$\GeV together with a veto on \PQb-tagged jets and additional leptons, as in the analysis selection.
We measure a data-to-MC correction factor of $0.95 \pm 0.06\stat$.
For CR$_{\mathrm{Z2}}$, the non-$\Z (\to \PGmp \PGmm)$ contributions, about 4\%, are treated as an uncertainty.
Adding the VBF topology selection defines CR$_{\mathrm{Z3}}$.
The ratio of CR$_{\mathrm{Z3}}$ to CR$_{\mathrm{Z2}}$ events in the data
gives the fraction of $\Z (\to \PGn\PAGn)$ + jets events passing
the VBF topology selection.
Table~\ref{table:results} details the contributions of the major backgrounds.

To determine the contribution of $\PW (\to \ell \nu)$ + jets background to the
analysis we use a similar procedure. We substitute the lepton veto with a one
muon requirement to obtain a $\PW (\to \mu \nu)$ plus two jets sample,
CR$_{\mathrm{W1}}$.  The data-to-MC correction factor for the sample is $0.97 \pm 0.01\stat$.
Treating the muon as undetected and requiring
$\mpt> 250$\GeV and the veto on \PQb-tagged jets and additional leptons as in the
analysis selection defines CR$_{\mathrm{W2}}$.
We measure a data-to-MC
correction factor of $0.80 \pm 0.04\stat$.
The control region CR$_{\mathrm{W3}}$ is obtained by adding the VBF topology selection.

From MC simulation we expect the fraction of events passing the lepton and jet vetoes and $\mpt$ selection that also satisfy our VBF topology selection to be the same
for the $\Z$ + jets and $\PW$ + jets events.
To increase the statistical precision we combine the two samples and obtain a prediction of $0.008 \pm 0.002\stat$.

The negligible contribution from QCD multijet production is checked using the number of events passing the analysis selection except the jet veto and $\abs{\Delta \phi ( \ptvecmiss, \text{jet}_2 )}$ requirement.
Nonmultijet background ($\Z/\PW$ + jets, $\ttbar$, diboson) is subtracted and the number of events is scaled by the efficiency to inefficiency ratios of the jet veto and $\abs{\Delta \phi ( \ptvecmiss, \text{jet}_2 )}$ requirements.
The two ratios are measured in low-$\mpt$ multijet-enriched data samples.
Other smaller background contributions ($\Z (\to \ell^{+} \ell^{-})$ + jets, $\ttbar$, diboson) are taken from simulation.

The dominant source of systematic uncertainty in the background
estimate for both $\Z (\to \PGn\PAGn)$ + jets and $\PW (\to \ell \nu)$
+ jets comes from the event yields found in the control regions.
The control sample statistics lead to an uncertainty in the data-to-MC correction factors
of 5--6\%, and 24\% on the fraction of events passing the VBF
topology selection.
Additional sources of systematic uncertainties due to trigger efficiency (5\%), background in the control regions (4--5\%), jet energy resolution and scale (3\%), and integrated luminosity measurement (3\%)~\cite{CMS:2013gfa} are incorporated.
The dominant source of systematic uncertainty in the signal expectation comes from the
modeling of the two jets in simulation, \ie the fraction of events
passing the VBF topology selection. We take the largest value of the observed difference
between data and MC of this fraction from the $\Z (\to \PGmp \PGmm)$
+ jets and $\PW (\to \mu \nu)$ + jets control regions, and their
uncertainties, as an estimate of the signal uncertainty. For the
uncertainty due to the choice of parton momentum distributions we follow
the PDF4LHC recommendations~\cite{Alekhin:2011sk, Botje:2011sn}, using
CTEQ6.6L, MRST2006, and NNPDF10~\cite{Nadolsky:2008zw,Martin:2007bv,
Ubiali:2008uk}. The dominant uncertainties that contribute to the
signal dijet mass shape include the $\mpt$ and jet energy scale
uncertainties.
The background dijet mass shape uncertainties, which vary between 7 and 42\%, are determined by comparing the differences in the predicted and measured dijet mass distributions in various low-$\mpt$ control regions for $\Z$ and $\PW$ + jets events.

\begin{table*}[tbh]
   \topcaption{Event yields predicted from MC in the control regions and observed in the signal region (SR). Only statistical uncertainties are shown. Dashes indicate cases where a background contribution is negligible. The units for the yields are given in the header row of the table.
}
  \centering
  \newcolumntype{x}{D{,}{\,\pm\,}{4}}
  \newcolumntype{X}{D{^}{^}{-1}}
  \renewcommand{\arraystretch}{1.2}
  \cmsTable{
      \begin{scotch}{l*{7}x}
          Sample
             & \multicolumn{1}{c}{CR$_{\mathrm{Z1}} \, (10^{3})$ }& \multicolumn{1}{c}{CR$_{\mathrm{Z2}}$} & \multicolumn{1}{c}{CR$_{\mathrm{Z3}}$}
             & \multicolumn{1}{c}{CR$_{\mathrm{W1}} \, (10^{4})$} & \multicolumn{1}{c}{CR$_{\mathrm{W2}} \, (10^{2})$} & \multicolumn{1}{c}{CR$_{\mathrm{W3}}$}
             & \multicolumn{1}{c}{SR} \\
          \hline
          $\PW (\to \ell \PGn)$ + jets
             & 0.10,0.02 & \multicolumn{1}{l}{$0.0^{+2.4}_{-0.0}$} & \multicolumn{1}{l}{$0.0^{+2.4}_{-0.0}$}
             & 6647,4 & 13.4,0.6 & 8.0,4.4
             & 43.6,10.3 \\
          $\Z (\to \PGn\PAGn)$ + jets & \multicolumn{1}{c}{\NA}  & \multicolumn{1}{c}{\NA}  & \multicolumn{1}{c}{\NA}
             & \multicolumn{1}{c}{\NA}  & \multicolumn{1}{c}{\NA}  & \multicolumn{1}{c}{\NA}  & 88.2,9.8 \\
	  $\Z (\to \ell^{+} \ell^{-})$ + jets
             & 5130,5 & 675,35 & 5.5,2.3
             & 594.9,0.4 & 0.12,0.04 & \multicolumn{1}{l}{$0.0^{+1.9}_{-0.0}$}
             & \multicolumn{1}{l}{$0.0^{+0.2}_{-0.0}$} \\
          $\ttbar$
             & 17.2,0.2 & 1.3,1.2 & \multicolumn{1}{l}{$0.0^{+0.7}_{-0.0}$}
             & 40.5,0.1 & 0.13,0.04 & \multicolumn{1}{l}{$0.0^{+0.7}_{-0.0}$}
             & \multicolumn{1}{l}{$0.0^{+0.7}_{-0.0}$} \\
          Diboson
             & 12.8,0.1 & 23.8,4.9 & \multicolumn{1}{l}{$0.02^{+0.25}_{-0.02}$}
             & 10.33,0.03 & 0.22,0.01 & \multicolumn{1}{l}{$0.07^{+0.34}_{-0.07}$}
             & \multicolumn{1}{l}{$0.4^{+0.7}_{-0.4}$} \\
          \hline
          $\Sigma$ MC
             & 5160,5 & 700,36 & 5.5,2.3
             & 7292,5 & 13.8,0.6 & 8.0,4.4
             & 132,14 \\
          Data & \multicolumn{1}{c}{5073} & \multicolumn{1}{c}{666} & \multicolumn{1}{c}{6}
             & \multicolumn{1}{c}{7075} & \multicolumn{1}{c}{11.1} & \multicolumn{1}{c}{9} & \multicolumn{1}{c}{118} \\
      \end{scotch}
}
\label{table:results}
\end{table*}

\begin{figure}[tbp]
\centering
      \includegraphics[width=0.96\linewidth]{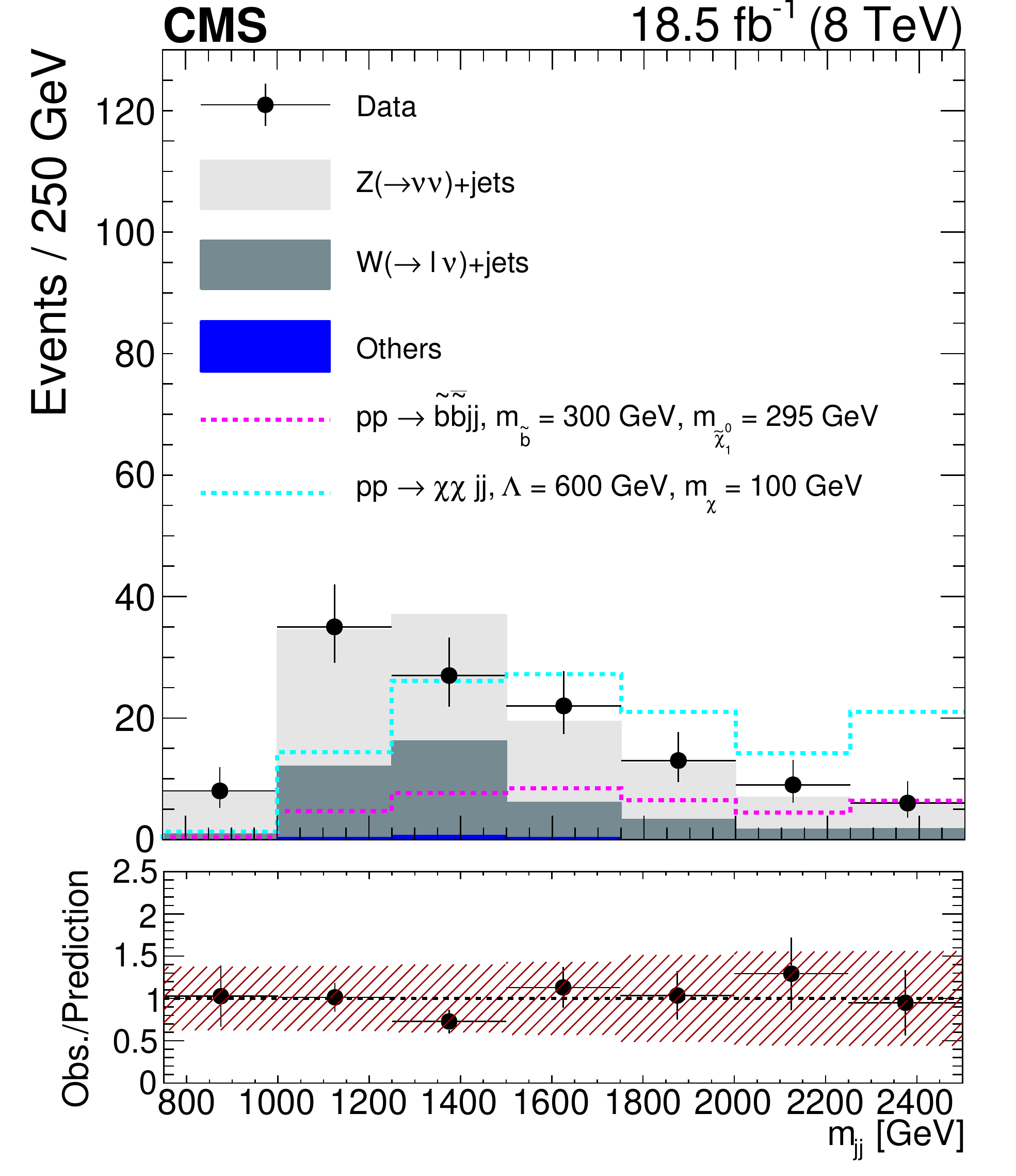}
      \caption{Dijet mass distribution of the data (dots), estimated background (stacked histograms), and signal samples (dashed lines) after the analysis selection.
	The last bin includes all events above 2250\GeV. The ratio plot (below) shows the yields in data divided by predicted yields for each bin.
         The shaded band in the ratio plot includes systematic and statistical uncertainties in the background prediction.}
      \label{fig:dijetmass}
\end{figure}

Figure~\ref{fig:dijetmass} shows the dijet mass distribution after
the analysis selection for the backgrounds and the two signal models.
Due to the harder scattering required for DM and bottom squark pair production, we expect a harder
dijet mass spectrum than for the SM backgrounds. We fit the dijet mass
distribution to calculate upper limits on the cross sections at a 95\%
confidence level (\CL) using the \CLs criterion~\cite{Read:2002hq,
Junk:1999kv} with the one-sided (LHC-style) profile likelihood ratio
as the test statistic.
Systematic uncertainties are represented by nuisance
parameters, assuming a gamma or
log-normal prior probability for normalization parameters and Gaussian priors for
dijet mass shape uncertainties.

\begin{figure*}[tbp]
\centering
\includegraphics[width=0.96\linewidth]{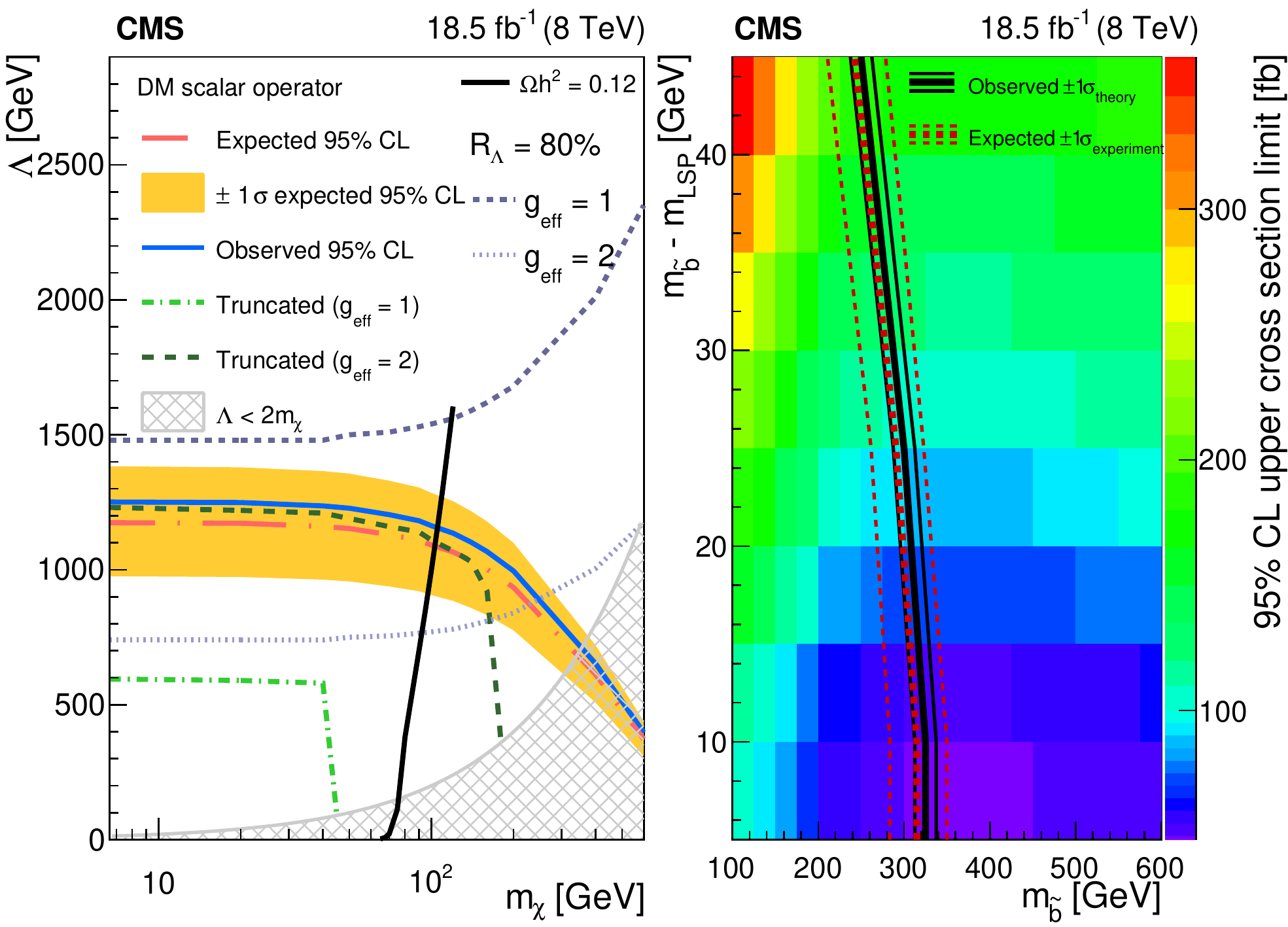}
\caption{(left) Contact interaction scale limit at 95\% CL as a function of the DM mass. The validity of the effective field theory is quantified by (i) ${R}_{\Lambda} = 80\%$ contours and (ii) truncated limits for different values of the effective coupling.
The DM relic abundance $\Omega h^{2}= 0.12$ is calculated as described in the text. (right) Bottom squark pair production 95\% CL upper cross section limit as a function of the bottom squark mass and the mass difference between the bottom squark and the LSP.
The observed (expected) cross section limit includes one standard deviation bands for the theoretical (experimental) uncertainty.
}
\label{fig:limits}
\end{figure*}

The result of the fit for the 95\% CL limit in the DM effective theory is given by the solid blue line in Figure~\ref{fig:limits} (left); values of ($m_{\chi}$, $\Lambda$) below the curve are excluded.
Although EFT is a good approximation in the regime of small momentum transfers, such as direct DM detection experiments, its validity needs to be quantified for LHC experiments, where interactions may occur with large momentum transfer.
For this purpose, an event in the MC signal sample is classified as having large momentum transfer if the center-of-mass energy of the DM pair ($E_{\text{cm}}^{\chi\chi}$) is larger than the mediator mass parameter of the EFT, $\mathcal{M} = \Lambda g_{\text{eff}}$.
In the EFT approach, each parameter point of $m_{\chi}$ and $\Lambda$ is classified as valid if the fraction of MC signal events ($R_{\Lambda}$) classified as not having large momentum transfer is 80\% or more.
Truncated limits are calculated by adding the requirement $E_{\text{cm}}^{\chi\chi} < \Lambda g_{\text{eff}}$ to the signal acceptance, following Refs.~\cite{Abercrombie:2015wmb, 2015arXiv151109375C}.
More signal events are removed in higher DM mass regions where
$R_{\Lambda}$ curves tend to go up and truncated limits go down. Figure~\ref{fig:limits} (left) shows curves corresponding to $R_{\Lambda} = 80\%$ and truncated limits for different values of $g_{\text{eff}}$, along with the DM
relic abundance $\Omega h^{2}= 0.12$ calculated using the \textsc{MadDM} program~\cite{Backovic201418}, assuming that DM pairs annihilate to electroweak boson pairs. The DM is more abundant than observed in the regions above or
left from the $\Omega h^{2}= 0.12$ line.

The observed cross section upper limit on bottom squark pair production in association with two partons ($\pt> 30\GeV$, $\abs{\Delta \eta} > 4.2$) is shown as a function of $m_{\PSQb}$ and its difference from $m_{\mathrm{LSP}}$ in Figure~\ref{fig:limits} (right).
The contours show observed and expected limits on the masses. The excluded mass values are taken at the intersection of the observed cross section limit with the theoretical cross section less one standard deviation of its uncertainty.

In summary, we have searched for new physics that results in large $\mpt$ and jets with a VBF topology. The data sample used corresponds to an integrated luminosity of 18.5\fbinv collected by the CMS detector in proton-proton
collisions at $\sqrt{s} = 8\TeV$. The low multijet background demonstrates the power of the VBF topology approach for DM and compressed mass spectrum SUSY searches. This is the first search for DM production through pure
electroweak VBF processes at a hadron collider. The production of DM via VBF with masses below 420\GeV is excluded at a 95\% confidence level for a chosen contact interaction scale $\Lambda = 600\GeV$, extending the reach by other
DM searches probing similar operators (e.g. $\Lambda$ exclusions up to $\approx$ 100 GeV for similar DM mass in~\cite{monoH1, monoH2, monoV, monoZ}).
Limits for different values of $\Lambda$ can be obtained by scaling the $\chi\chi
\mathrm{j j}$ cross section, which is proportional to $1 / \Lambda^{2}$.
For a nearly mass-degenerate bottom squark and LSP this analysis sets the most stringent limits reported to date, excluding scalar bottom quarks up to masses of 315\GeV at a 95\% confidence level.

\begin{acknowledgments}
We congratulate our colleagues in the CERN accelerator departments for the excellent performance of the LHC and thank the technical and administrative staffs at CERN and at other CMS institutes for their contributions to the success of the CMS effort. In addition, we gratefully acknowledge the computing centers and personnel of the Worldwide LHC Computing Grid for delivering so effectively the computing infrastructure essential to our analyses. Finally, we acknowledge the enduring support for the construction and operation of the LHC and the CMS detector provided by the following funding agencies: BMWFW and FWF (Austria); FNRS and FWO (Belgium); CNPq, CAPES, FAPERJ, and FAPESP (Brazil); MES (Bulgaria); CERN; CAS, MoST, and NSFC (China); COLCIENCIAS (Colombia); MSES and CSF (Croatia); RPF (Cyprus); MoER, ERC IUT and ERDF (Estonia); Academy of Finland, MEC, and HIP (Finland); CEA and CNRS/IN2P3 (France); BMBF, DFG, and HGF (Germany); GSRT (Greece); OTKA and NIH (Hungary); DAE and DST (India); IPM (Iran); SFI (Ireland); INFN (Italy); MSIP and NRF (Republic of Korea); LAS (Lithuania); MOE and UM (Malaysia); BUAP, CINVESTAV, CONACYT, LNS, SEP, and UASLP-FAI (Mexico); MBIE (New Zealand); PAEC (Pakistan); MSHE and NSC (Poland); FCT (Portugal); JINR (Dubna); MON, RosAtom, RAS and RFBR (Russia); MESTD (Serbia); SEIDI and CPAN (Spain); Swiss Funding Agencies (Switzerland); MST (Taipei); ThEPCenter, IPST, STAR and NSTDA (Thailand); TUBITAK and TAEK (Turkey); NASU and SFFR (Ukraine); STFC (United Kingdom); DOE and NSF (USA).

Individuals have received support from the Marie-Curie program and the European Research Council and EPLANET (European Union); the Leventis Foundation; the A. P. Sloan Foundation; the Alexander von Humboldt Foundation; the Belgian Federal Science Policy Office; the Fonds pour la Formation \`a la Recherche dans l'Industrie et dans l'Agriculture (FRIA-Belgium); the Agentschap voor Innovatie door Wetenschap en Technologie (IWT-Belgium); the Ministry of Education, Youth and Sports (MEYS) of the Czech Republic; the Council of Science and Industrial Research, India; the HOMING PLUS program of the Foundation for Polish Science, cofinanced from European Union, Regional Development Fund; the Mobility Plus program of the Ministry of Science and Higher Education (Poland); the OPUS program of the National Science Center (Poland); the Thalis and Aristeia programs cofinanced by EU-ESF and the Greek NSRF; the National Priorities Research Program by Qatar National Research Fund; the Programa Clar\'in-COFUND del Principado de Asturias; the Rachadapisek Sompot Fund for Postdoctoral Fellowship, Chulalongkorn University (Thailand); the Chulalongkorn Academic into Its 2nd Century Project Advancement Project (Thailand); and the Welch Foundation, contract C-1845.
\end{acknowledgments}
\bibliography{auto_generated}
\cleardoublepage \appendix\section{The CMS Collaboration \label{app:collab}}\begin{sloppypar}\hyphenpenalty=5000\widowpenalty=500\clubpenalty=5000\textbf{Yerevan Physics Institute,  Yerevan,  Armenia}\\*[0pt]
V.~Khachatryan, A.M.~Sirunyan, A.~Tumasyan
\vskip\cmsinstskip
\textbf{Institut f\"{u}r Hochenergiephysik der OeAW,  Wien,  Austria}\\*[0pt]
W.~Adam, E.~Asilar, T.~Bergauer, J.~Brandstetter, E.~Brondolin, M.~Dragicevic, J.~Er\"{o}, M.~Flechl, M.~Friedl, R.~Fr\"{u}hwirth\cmsAuthorMark{1}, V.M.~Ghete, C.~Hartl, N.~H\"{o}rmann, J.~Hrubec, M.~Jeitler\cmsAuthorMark{1}, A.~K\"{o}nig, I.~Kr\"{a}tschmer, D.~Liko, T.~Matsushita, I.~Mikulec, D.~Rabady, N.~Rad, B.~Rahbaran, H.~Rohringer, J.~Schieck\cmsAuthorMark{1}, J.~Strauss, W.~Treberer-Treberspurg, W.~Waltenberger, C.-E.~Wulz\cmsAuthorMark{1}
% ...
\vskip\cmsinstskip
% ...
1:~~Also at Vienna University of Technology, Vienna, Austria\\
% ...

\end{sloppypar}
\end{document}